\documentclass[a4paper, amsfonts, amssymb, amsmath, reprint, showkeys, nofootinbib, twoside]{revtex4-1}
\usepackage[english]{babel}
\usepackage[utf8]{inputenc}
\date{April 04, 2021}

\begin{document}
\title{The effects of $q$-statistics on cosmology}
\author{M. Senay \\Naval Academy, National Defence University, 34940, Istanbul, Turkey\\ mustafasenay86@hotmail.com}
\begin{abstract}
   Considering Verlinde's entropic gravity proposal, we focus the effects of fermionic $q$ deformation on the Einstein's field equations and Friedmann equations. For this purpose, we represent the thermodynamical properties of the deformed fermion gas model in two-dimensional space. To describe the behavior of the quantum black holes, deformed Einstein field equations are derived with the help of deformed entropy function. Moreover, deformed Friedmann equations are investigated by using Friedmann-Robertson-Walker (FRW) metric. We also present the effective ones of the energy density, the pressure, and the equation of state for the dark energy. Lastly, we derive an analytical expression of the effective density parameter of dark energy.
\end{abstract}
\maketitle
\section{Introduction} 
With the discovery of thermodynamic properties of black holes, there has been found out a deep connection between gravity and thermodynamics. In 1971, Hawking \cite{Hawking1971} demonstrated that the area of black hole event horizon cannot be reduced by any process and when two black holes merged, the generated new black hole area cannot be smaller than the sum of the initial black hole areas. In 1973, Bekeinstein  \cite{Bekeinstein1973} showed that the surface of the event horizon of the black hole and its entropy were related with each other. In 1995, Jacobson  \cite{Jacobson1995} indicated that Eisntein's field equations can be derived by using the first law of thermodynamics.\par
In 2011, Verlinde  \cite{Verlinde2011} suggested a  new remarkable perspective on the relation between gravity and thermodynamics. According to Verlinde's suggestion, gravity can be explained as an entropic force when taken into holographic viewpoint of gravity. Using the entropic force together with the holographic principle and the equipartition law of energy, he derived Newton's law of gravitation, and Einstein's field equations. \par
The holographic approach on gravity has enabled to conduct many studies in the literature \cite{Sheykhi2012,Moradpour2015,Moradpour2018,Sefiedgar2017,Feng2016,Abreu2018,Abbasi2020,Saridakis2020}. For example, in the Ref \cite{Sheykhi2012}, the equipartition law of energy theorem was modified with Debye function and MOND theory and Einstein's field equations were derived under Debye entropic gravity scenario. In the Ref \cite{Saridakis2020}, modifed Friedmann equations were obtained by using Barrow entropy instead of usual Bekeinstein-Hawking entropy. \par
On the other hand, the mass of a charged black hole decreases via Hawking radiation until it reaches a minimum mass proportional to it's charge \cite{Strominger1993}. The black hole having a minimum mass is assumed as an extremal black hole. The extremal black holes can be used to find the solution to the quantum puzzles of black holes. For instance, Strominger \cite{Strominger1993} considered the scattering of two extremal black holes for solving the quantum black hole puzzles. He looked to answer the question of whether do extremal black holes scatter as bosons, fermions or something else? According to the results from his study, the extremal black holes obey the deformed statistics instead of usual statistics. Hence, they can be considered as deformed bosons or fermions. In recent years, several authors have intensively examined both the quantum mechanical and thermodynamical properties of such deformed bosons and fermions \cite{Lavagno2002, Gavrilik2011, Algin2016a}, and their applications on different systems \cite{Dil2015,Dil2017,Senay2018,Kibaroglu2019}.\par
In the view of the above motivations, in this study, based on Strominger's idea we specially consider extremal black holes as $q$-deformed fermions. Thermodynamical properties of these $q$-deformed fermions were investigated in Ref. \cite{Algin2016b}. The present paper is aimed to investigate the effects of fermionic \emph{q}-deformation on the Einstein's field equations and Friedmann equations. Hence, some thermodynamical properties of deformed fermion gas model in two dimensional space is briefly introduced in Section II. With the help of deformed entropy function, deformed Einstein's field equations are found in Section III. The deformed Friedman equations are derived  in Section IV. The cosmological implications are given in Section V. The conclusions are given in the last section.
\section{Deformed fermion statistics in two dimensions}
The non-symmetric deformed fermion oscillators algebra is defined in terms of the creation operator $f^{*}$ and annihilation operator $f$ in the following form \cite{Parthasarathy1991,Viswanathan1992,Chaichian1993,Algin2016b},
\begin{equation}
ff^{*}+qf^{*}f=1,
\end{equation}
\begin{equation}
[\hat{N},f^{*}]=f^{*},\,\,\,\,[\hat{N},f]=-f,
\end{equation}
where $\hat{N}$ is the fermionic number operator and $\emph{q}$ is the real positive deformation parameter which takes values in the interval $0<\emph{q}<1$. The basic deformed quantum number is given as
\begin{equation}
    [n]=\frac{1-(-1)^nq^n}{1+q}.
\end{equation}
To investigate the thermostatistical properties of the deformed fermion gas model, the fermionic Jackson derivative operator is used instead of the standard derivative operator 
\begin{equation}
    D_{x}^{q}f(x)=\frac{1}{x}\left[\frac{f(x)-f(-qx)}{1+q}\right],
\end{equation}
for any function $f(x)$. The mean occupation number of the deformed fermion gas model is expressed as
\begin{equation}
    n_i=\frac{1}{|lnq|}\left|ln\left(\frac{|1-ze^{\beta\epsilon_i}|}{1+qze^{\beta\epsilon_i}}\right)\right|,
\end{equation}
where $\epsilon_i$ is the kinetic energy of a particle in the one-particle energy state, $\beta=1/k_BT$, $k_B$ is the Boltzmann constant, $T$ is the temperature of the system and the fugacity $z=exp(\mu/k_BT)$ has the standard form. \par
Using the Eq. (5), the total number of particles and the energy of the system can be, respectively, expressed as
\begin{equation}
    N=\sum_{i}\frac{1}{|lnq|}\left|ln\left(\frac{|1-ze^{\beta\epsilon_i}|}{1+qze^{\beta\epsilon_i}}\right)\right|,
\end{equation}
\begin{equation}
    U=\sum_{i}\frac{\epsilon_i}{|lnq|}\left|ln\left(\frac{|1-ze^{\beta\epsilon_i}|}{1+qze^{\beta\epsilon_i}}\right)\right|.
\end{equation}
For a large volume and a large number of particles, we can replace the sums over states by integrals. Using the density of state in two dimensions $g(\epsilon)=(2{\pi}{mA}/h^2)$, one can easily reach the following relations
\begin{equation}
    \frac{N}{A}=\frac{1}{\lambda^2}f_1(z,q),
\end{equation}
\begin{equation}
    \frac{U}{A}=\frac{1}{\lambda^2}k_BTf_2(z,q),
\end{equation}
where $\lambda=h/(2{\pi}{mk_BT})^{1/2}$ is the thermal wavelength and the generalized Fermi-Dirac function $f_n(z,q)$ is defined as
\begin{eqnarray}
     f_n(z,q) & = & \frac{1}{\Gamma (n)}\int_{0}^{\infty}{\frac{x^{n-1}}{|lnq|}}\left|ln\left(\frac{|1-ze^{-x}|}{1+qze^{-x}}\right)\right|dx\nonumber \\
 & = & \frac{1}{|lnq|}\left[\sum_{l=1}^{\infty}\frac{(zq)^l}{l^{n+1}}-\sum_{l=1}^{\infty}\frac{z^l}{l^{n+1}}\right],
\end{eqnarray}
where $x=\beta\epsilon$. From the thermodynamic relation $F=\mu{N}-PA$, the Helmholtz free energy in two dimensions can be found as
\begin{equation}
    F=\frac{k_BTA}{\lambda^2}\left[{f_1(z,q)lnz-f_2(z,q)}\right].
\end{equation}
The entropy function of the model can be obtained by the relation $S=(U-F)/T$ as
\begin{equation}
    S=\frac{k_BA}{\lambda^2}\left[{2f_2(z,q)-f_1(z,q)lnz}\right].
\end{equation}
If we take the one particle kinetic energy $E=k_BT$ the Eq. (12) can be re-expressed as
\begin{equation}
    S=\frac{2{\pi}mA}{h^2T}E^2F(z,q),
\end{equation}
where
\begin{equation}
    F(z,q)=2f_2(z,q)-f_1(z,q)lnz.
\end{equation}\par
\section{The Einstein field equations due to deformed entropic gravity }
In this section, we investigate the fermionic $q$-deformation effect on Einstein's field equations under Verlinde's approach. Acoording to Verlinde's proposal \cite{Verlinde2011}, when a test particle approaches a holographic screen, the entropic force of the system is expressed as
\begin{equation}
    \mathcal{F}=T\frac{\Delta{S}}{\Delta{x}},
\end{equation}
where $\Delta{S}$ is the change of the entropy on holographic screen and $\Delta{x}$ is the displacement of the particle from holographic screen. The holographic screen can be thought as maximal storage space for information. The total number of bits $N$ is assumed to be proportional to the area $A$ of the holographic screen and it is given as
\begin{equation}
    N=\frac{Ac^3}{2G\hbar}.
\end{equation}
When the entropic force is equal to the force increasing the entropy, the total entropy of the system remains constant. Therefore, the variation of the entropy goes to zero when the system has statistical equilibrium
\begin{equation}
    \frac{d}{dx^a}S(E,x^a)=0,
\end{equation}
and it can be re-expressed as
\begin{equation}
    \frac{\partial{S}}{\partial{E}}{\frac{\partial{E}}{\partial{x^a}}+\frac{\partial{S}}{\partial{x^a}}}=0,
\end{equation}
where $\frac{\partial{E}}{\partial{x^a}}=-F_a$ and $\frac{\partial{S}}{\partial{x^a}}=\nabla_a{S}$. Substituting Eq. (13) into Eq. (18), the deformed temperature can be derived as
\begin{equation}
   T=\frac{mAE}{\pi{h}}F(z,q)e^\phi{N^a}{\nabla_a{\phi}},
\end{equation}
where we have used $F_a=me^\phi{\nabla_a}\phi$ and $\nabla_a{S}=-2\pi{mN_a}/\hbar$ in the last equation. Also, $e^\phi$ indicates the redshift factor and $N^a$ is the unit outward pointing vector. The above temperature equation is re-expressed as
\begin{equation}
    T=2\tilde{\alpha}({z,q})T_U
\end{equation}
where $T_U=\frac{\hbar}{2\pi}e^\phi{N^a}{\nabla_a{\phi}}$ is the Unruh temperature \cite{Unruh1976,Verlinde2011} and $\tilde{\alpha}(z,q)$ carries the information of deformed fermion system and defined as
\begin{equation}
    \tilde{\alpha}(z,q)=\frac{2\pi{mAE}}{h^2}F(z,q).
\end{equation}
Using the relation between mass-energy $M=N{T}/2$, the total mass can be written as
\begin{equation}
   M=\frac{\tilde{\alpha}(z,q)}{4\pi{G}}\int_{\mathcal{S}}e^\phi{\nabla{\phi}}dA,
\end{equation}
where $\mathcal{S}$ is the holographic screen. The integral on the right hand side relates with the modified Komar mass. Hence, the Eq. (22) can be considered as the modified Gauss's law in general relativity. Using the Stokes theorem and following same procedure in Ref. \cite{Senay2018, Kibaroglu2019}, the Eq. (22) can be written in terms of the Killing vector $\xi^a$ and Ricci curvature tensor $R_{ab}$ as
\begin{equation}
   M=\frac{\tilde{\alpha}(z,q)}{4\pi{G}}\int_{V}R_{ab}n^a\xi^bdV,
\end{equation}
where $V$ is a spacelike hypersurface demonstrating the space volume and $n_a$ is the unit vector normal to $V$. Moreover, the Komar mass can be expressed in terms of the stress-energy tensor $T_{ab}$ \cite{Wald1984} as
\begin{equation}
   \mathcal{M}=2\int_{V}\left(T_{ab}-\frac{1}{2}g_{ab}T+\frac{\Lambda}{8\pi{G}}g_{ab}\right){n^a\xi^b}dV,
\end{equation}
where $g_{ab}$ is metric tensor. Equating Eq. (23) and Eq. (24), we obtain
\begin{equation}
   \tilde{\alpha}(z,q){R_{ab}}=8\pi{G}\left(T_{ab}-\frac{1}{2}g_{ab}T+\frac{\Lambda}{8\pi{G}}g_{ab}\right).
\end{equation}
If we take the trace of the last equation, we reach
\begin{equation}
   R_{ab}-{\frac{1}{2}g_{ab}R+{\frac{\Lambda}{\tilde{\alpha}(z,q)}}g_{ab}=\frac{8\pi{G}}{\tilde{\alpha}(z,q)}T_{ab}},
\end{equation}
Thus, we get a modification of Einstein's field equations resulting from deformed fermion theory under Verlinde's entropic gravity scenario.
\section{Friedmann equation due to deformed entropic gravity}
In this section, we want to derive modifed version of Friedmann equations. In the homogeneous and isotropic space time, the FRW universe is given by the line element
\begin{equation}
   ds^2=c^2dt^2-a^2(t)\left(\frac{dr^2}{1-kr^2}+r^2d\Omega^2\right),
\end{equation}
where $a(t)$ is the scale fator and $k$ is constant relating curvature of the universe with $k=0,1,-1$ corresponding to flat, closed, and open universes, respectively. The dynamical apparent horizon for the FRW univers can be defined as
\begin{equation}
    \tilde{r}=ar=\frac{1}{\sqrt{H^2+k/a^2}},
\end{equation}
where $H=\dot{a}/a$ is the Hubble parameter and the dots represents the time derivative. Now, assuming that the matter source in the FRW universe can be expressed as a perfect fluid with the following stress-energy tensor
\begin{equation}
   T_{ab}=\left(\rho+p\right)u_au_b-pg_{ab},
\end{equation}
where $\rho$ is the mass-energy density, $p$ is the pressure of the fluid, and $u_a$ is the fluid's four vector. The conservation law of energy-momentum leads to the following continuity equation
\begin{equation}
   \dot{\rho}+3H\left(\rho+p\right)=0.
\end{equation}
Now, we consider a compact spatial region $V=\frac{4}{3}\pi\tilde{r}^3$ with a compact boundary $\mathcal{S}=4\pi\tilde{r}^2$. The acceleration of radial observe is given as
\begin{equation}
    a_r=-\ddot{a}r,
\end{equation}
and it leads to take the following form for Unruh temperature
\begin{equation}
    T_U=\frac{\hbar{a_r}}{2\pi}.
\end{equation}
The total physical mass $M$ inside the volume $V$ can be defined as
\begin{equation}
    M=\int_V{dV(T_{ab}u^au^b)}=\frac{4}{3}\pi\tilde{r}^3\rho.
\end{equation}
Using the Eqs. (16), (20), (31)-(33), one can easily find
\begin{equation}
    \frac{\ddot{a}}{a}=-\frac{4\pi{G}}{3\tilde{\alpha}(z,q)}\rho
\end{equation}
which is modified version of the dynamical equations for the Newtonian cosmology in the deformed case. Now, we want to derive the Friedmann equations of FRW universe. For this purpose, we need to use the active gravitational mass $\mathcal{M}$ instead of the total mass $M$, since the active gravitational mass produces the acceleration in the dynamical background. In our context, the active gravitational mass can defined as
\begin{equation}
    \mathcal{M}=\left(\rho+3p+\frac{\Lambda}{4\pi{G}}\right)V.
\end{equation}
Replacing the total mass $M$ with active gravitational mass $\mathcal{M}$ in the Eq. (33), we reach,
\begin{equation}
    \frac{\ddot{a}}{a}=-\frac{4\pi{G}}{3\tilde{\alpha}(z,q)}(\rho+3p)-\frac{\Lambda}{3\tilde{\alpha}(z,q)},
\end{equation}
which is the deformed acceleration equation for the dynamical evolution of the FRW universe. Using the Eq. (30) and multiplying both sides of Eq. (36) with $a\dot{a}$ and then integrating it, we have
\begin{equation}
    H^2+\frac{k}{a^2}=\frac{8\pi{G}}{3}\left(\frac{\rho}{\tilde{\alpha}(z,q)}-\frac{\Lambda}{8\pi{G}\tilde{\alpha}(z,q)}\right).
\end{equation}
For simplicity we focus on the flat case $k=0$, the last equation can re-express as
\begin{equation}
     H^2=\frac{8\pi{G}}{3}(\rho+\rho_{DE}),
\end{equation}
where $\rho_{DE}$ is the effective energy density of the dark energy
\begin{equation}
    \rho_{DE}=\left(\frac{1}{\tilde{\alpha}(z,q)}-1\right)\rho-\frac{\Lambda}{8\pi{G}\tilde{\alpha}(z,q)}.
\end{equation}
Moreover, using the relation $\dot{H}=\ddot{a}/a-H^2$ together with Eqs. (36)-(39), one can easily obtain the following expression
\begin{equation}
    \dot{H}=-4\pi{G}(\rho+p+\rho_{DE}+p_{DE}),
\end{equation}
where $p_{DE}$ is the effective pressure of the dark energy
\begin{equation}
    p_{DE}=\left(\frac{1}{\tilde{\alpha}(z,q)}-1\right)p+\frac{\Lambda}{8\pi{G}\tilde{\alpha}(z,q)}.
\end{equation}
Moreover, the effective equation of state can be easily found in the following form:
\begin{equation}
    w_{DE}=\frac{p_{DE}}{\rho_{DE}}=\frac{\left(\frac{1}{\tilde{\alpha}(z,q)}-1\right)p+\frac{\Lambda}{8\pi{G}\tilde{\alpha}(z,q)}}{\left(\frac{1}{\tilde{\alpha}(z,q)}-1\right)\rho-\frac{\Lambda}{8\pi{G}\tilde{\alpha}(z,q)}}.
\end{equation}
Before we close this section, we want to remark that we considered a static background by following Verlinde's way. In order to obtain Einstein's equation, he uses a timelike Killing vector which exists for static or stationary spacetime \cite{Cai2010}. In case of the FRW spacetime, this can be possible only if the metric Eq. (27) reduces to the de Sitter or Minkowski spacetime \cite{Tower2014}. By considering this assumption, we have obtained the modified version of the dynamical equations governing the evolution of the FRW universe.
\section{Cosmological Results}
In this section, we briefly investigate the comological evolution in the scenario of $q$-deformed dark energy model. Now we consider the case of dust matter (${p}\approx{0}$) in which case Eq. (30) leads to $\rho=\rho_{0}a^{-3}$ \cite{Roos2015}, where $\rho_{0}$ is the current value of the dust density. Furthermore, the density parameters of the matter and the effective dark energy can be defined in the following form, respectively
\begin{equation}
    \Omega_m=\frac{8\pi{G}}{3H^2}\rho,
\end{equation}
\begin{equation}
    \Omega_m=\frac{8\pi{G}}{3H^2}\rho_{DE}.
\end{equation}
From Eq. (43) we reach $\Omega_m=(\Omega_{m0}{H^2_0})/(a^{3}H^{2}) $ and using the definition
\begin{equation}
    \Omega_{m}+\Omega_{DE}=1,
\end{equation}
we find
\begin{equation}
    H=\frac{\sqrt{\Omega_{m0}}{H_0}}{\sqrt{a^3(1-\Omega_{DE})}}.
\end{equation}
Differentiating last equation and using redsift $z_R=a^{-1}-1$, we obtain
\begin{equation}
    \dot{H}=-\frac{H^2}{2(1-\Omega_{DE})}\left(3(1-\Omega_{DE})+(1+z_R)\Omega^\prime_{DE}\right)
\end{equation}
where the prime represents $z_R$-derivative. Inserting Eq. (39) into Eq. (44) and using Eq. (46), we can easily reach
\begin{equation}
    \Omega_{DE}=1-\frac{1}{1+\frac{8\pi{G}}{3\Omega_{m0}H_0^2(1+z_R)^3}\left[\left(\frac{1}{\tilde{\alpha}(z,q)}-1\right)\rho-\frac{\Lambda}{8\pi{G}\tilde{\alpha}(z,q)}\right]},
\end{equation}
which is the analytical solution of the dark energy density parameter for the dust matter case in a flat universe.
\section{Conlusion}
In this paper, we introduced some of the high temperature thermodynamical properties of $q$-deformed fermion gas model in two-dimensional space, and particularly discussed its effects on the Einstein field equations, Friedmann equations and dark energy. For this aim, we considered quantum black holes as a deformed fermion particles proposed by Strominger\cite{Strominger1993} and used entropic gravity assumption proposed by Verlinde \cite{Verlinde2011}.\par
By the help of the $q$-deformed entropy function in Eq. (12), we obtained $q$-deformed temperature function in Eq. (19). Using it in the mass-energy relation and Komar mass definition, we derived $q$-deformed Einstein field equations in Eq. (26). The factor $\tilde{\alpha}(z,q)$ in the Eq. (26) carries the information of $q$-deformed fermion gas model. This factor does not reduce to unity in the limit $q\rightarrow1$. Therefore, the $q$-deformed Einstein field equations do not reduce to standard ones when considered $q\rightarrow1$.\par
Moreover, it can be seen in Eq. (26) that the field equations remain intant in their original form while the $q$-deformation brings a certain kind of re-scaling to the two constant, namely the Newton's gravitational constant and cosmological constant. For a more novel approach, the field equations ought to get correction to the curvature or the matter sector rather a mere re-scaling of certain constants. Although $q$ is a free parameter which varies between 0 and 1, both $G$ and $\Lambda$ might be varying constant. The $q$-formation to field equations would not effect any physical processes under consideration. \par
Considering FRW metric, we could obtain $q$-deformed version of Friedmann equations in Eqs. (36) and (37). These equations do not coincide with $\Lambda{CDM}$ paradigm due to the factor $\tilde{\alpha}(z,q)$. However, these may be considered as the generalized version of $\Lambda{CDM}$ model. Also, taking into account flat case $k=0$, we determined the effective energy density, the effective pressure, and the effective equation of state for the dark energy sector. Moreover, supposing  a background filled by a pressureless source, we derived analytical expression of the effective density parameter of dark energy in Eq. (48). \par
Consequently, the $q$-deformed fermion theory plays a crucial role on the investigation of cosmological evolution under gravity-thermodynamic connection. We hope that the results obtained this study may be used to understand the properties of the early universe and dark energy.\par
The investigation of the present $q$-deformed fermion model on the Einstein field equations and cosmology in the low-temperature limit is one of the open problems to study in the near future.


\begin{thebibliography}{0}%
\makeatletter
\providecommand \@ifxundefined [1]{%
 \@ifx{#1\undefined}
}%
\providecommand \@ifnum [1]{%
 \ifnum #1\expandafter \@firstoftwo
 \else \expandafter \@secondoftwo
 \fi
}%
\providecommand \@ifx [1]{%
 \ifx #1\expandafter \@firstoftwo
 \else \expandafter \@secondoftwo
 \fi
}%
\providecommand \natexlab [1]{#1}%
\providecommand \enquote  [1]{``#1''}%
\providecommand \bibnamefont  [1]{#1}%
\providecommand \bibfnamefont [1]{#1}%
\providecommand \citenamefont [1]{#1}%
\providecommand \href@noop [0]{\@secondoftwo}%
\providecommand \href [0]{\begingroup \@sanitize@url \@href}%
\providecommand \@href[1]{\@@startlink{#1}\@@href}%
\providecommand \@@href[1]{\endgroup#1\@@endlink}%
\providecommand \@sanitize@url [0]{\catcode `\\12\catcode `\$12\catcode
  `\&12\catcode `\#12\catcode `\^12\catcode `\_12\catcode `\%12\relax}%
\providecommand \@@startlink[1]{}%
\providecommand \@@endlink[0]{}%
\providecommand \url  [0]{\begingroup\@sanitize@url \@url }%
\providecommand \@url [1]{\endgroup\@href {#1}{\urlprefix }}%
\providecommand \urlprefix  [0]{URL }%
\providecommand \Eprint [0]{\href }%
\providecommand \doibase [0]{http://dx.doi.org/}%
\providecommand \selectlanguage [0]{\@gobble}%
\providecommand \bibinfo  [0]{\@secondoftwo}%
\providecommand \bibfield  [0]{\@secondoftwo}%
\providecommand \translation [1]{[#1]}%
\providecommand \BibitemOpen [0]{}%
\providecommand \bibitemStop [0]{}%
\providecommand \bibitemNoStop [0]{.\EOS\space}%
\providecommand \EOS [0]{\spacefactor3000\relax}%
\providecommand \BibitemShut  [1]{\csname bibitem#1\endcsname}%
\let\auto@bib@innerbib\@empty
\end{thebibliography}%


\begin{thebibliography}{10}
\bibitem{Hawking1971} S. W. Hawking, Phys. Rev. Lett., 26 (1971), pp. 1344-1346.
\bibitem{Bekeinstein1973} J. D. Bekeinstein, Phys. Rev. D, 7 (1973), pp. 2333-2346.
\bibitem{Jacobson1995} T. Jacobson, Phys. Rev. Lett., 75 (1995), pp. 1260-1263.
\bibitem{Verlinde2011} E. Verlinde, JHEP, 2011 (2011), Article 29.
\bibitem{Sheykhi2012} A. Sheykhi, K. R. Sarab, JCAP, 2012 (2012), Article 012.
\bibitem{Moradpour2015} H. Moradpour, A. Sheykhi, Int. J. Theor. Phys., 55 (2015), pp. 1145-1155.
\bibitem{Moradpour2018} H. Moradpour, A. Sheykhi, C. Corda, I. G. Salako, Phys. Lett. B, 783 (2018), pp. 82-85.
\bibitem{Sefiedgar2017} A. S. Sefiedgar, EPL, 117 (2017), Article 69001.
\bibitem{Feng2016} Z. W. Feng, S. Z. Yang, H. L. Li, X. T. Zu, Adv. High Energy Phys., 2016 (2016), Article 2341879.
\bibitem{Abreu2018} E. M. C. Abreu, J. A. Neto, A. C. R. Mendes, A. Bonilla, R.  M. de Paula, EPL, 124 (2018), Article 3.
\bibitem{Abbasi2020} K. Abbasi, S. Gharaati, Adv. High Energy Phys., 2020 (2020), Article 9362575.
\bibitem{Saridakis2020} E. N. Saridakis, JCAP, 2020 (2020), Article 07.
\bibitem{Strominger1993} A. Strominger, Phys. Rev. Lett., 71 (1993), pp. 3397-3400.
\bibitem{Lavagno2002} A. Lavagno and P. Narayana Swamy, Phys. Rev. E, 65 (2002), Article 036101.
\bibitem{Gavrilik2011} A. M. Gavrilik, I. I. Kachurik and Y. U. Mishchenko, J. Phys. A: Math. Theor., 44 (2011), Article 475303.
\bibitem{Algin2016a} A. Algin, M. Senay, Physica A, 447 (2016), pp. 232-246.
\bibitem{Dil2015} E. Dil, Can. J. Phys., 93 (2015), pp. 1274-1278.
\bibitem{Dil2017} E. Dil,  Int. J. Mod. Phys. A, 32 (2017), Article 205042.
\bibitem{Senay2018} M. Senay, S. Kibaroğlu, Int. J. Mod. Phys. A, 33 (2018), Article 18502128.
\bibitem{Kibaroglu2019} S. Kibaroğlu, M. Senay, Mod. Phys. Lett. A, 34 (2019), Article 1950249.
\bibitem{Algin2016b} A. Algin, M. Senay, J. Phys.: Conf. Ser., 766 (2016), Article 012008.
\bibitem{Parthasarathy1991} R. Parthasarathy, K. S. Viswanathan, J. Phys. A: Math. Gen., 24, (1991), pp. 613-617.
\bibitem{Viswanathan1992} K. S. Viswanathan, R. Parthasarathy, R. Jagannathan, J. Phys. A: Math. Gen., 25 (1992), pp. L335-L339.
\bibitem{Chaichian1993} M. Chaichian, R. G. Felipe, C. Montonen, J. Phys. A: Math. Gen., 26 (1993), pp. 4017-4034.
\bibitem{Wald1984} R. M. Wald, General Relativity, University of Chicago Press, 1984.
\bibitem{Unruh1976} W. G. Unruh, Phys. Rev. D, 14 (1976), pp. 870-892.
\bibitem{Cai2010} R. G. Cai, L. M. Cao, N. Ohta, Phys. Rev. D, 81 (2010), Article 061501.
\bibitem{Tower2014} W. Tower, Sci. Chin Phys. Math. Astro., 57 (2014), pp. 1623-1629.
\bibitem{Roos2015} M. Roos, Introduction to Cosmology, John Wiley and Sons, 2015.
\end{thebibliography}
\end{document}